\documentclass[10pt,letterpaper]{article}
\usepackage[top=0.85in,left=2.75in,footskip=0.75in]{geometry}
\usepackage{amsmath,amssymb}
\usepackage{changepage}
\usepackage[utf8x]{inputenc}
\usepackage{textcomp,marvosym}
\usepackage{cite}
\usepackage{nameref,hyperref}
\usepackage[right]{lineno}
\usepackage{microtype}
\DisableLigatures[f]{encoding = *, family = * }
\usepackage[table]{xcolor}
\usepackage{array}
\usepackage{float}
\newcolumntype{+}{!{\vrule width 2pt}}
\usepackage{comment}

\newlength\savedwidth

\raggedright
\usepackage[aboveskip=1pt,labelfont=bf,labelsep=period,justification=raggedright,singlelinecheck=off]{caption}

\makeatletter
\renewcommand{\@biblabel}[1]{\quad#1.}
\makeatother

\usepackage{lastpage,fancyhdr,graphicx}
\usepackage{epstopdf}
\fancyheadoffset[L]{2.25in}
\fancyfootoffset[L]{2.25in}
\begin{document}
\vspace*{0.2in}

\begin{flushleft}
{\Large
\textbf\newline{Efficient, high-performance semantic segmentation using multi-scale feature extraction}
}
\newline
\\
Moritz Knolle\textsuperscript{\Yinyang1,2},
Georgios Kaissis\textsuperscript{\Yinyang1,2,3,4},
Friederike Jungmann\textsuperscript{1},
Sebastian Ziegelmayer\textsuperscript{1},
Daniel Sasse\textsuperscript{1},
Marcus Makowski\textsuperscript{1},
Daniel Rueckert\textsuperscript{2,4},
Rickmer Braren\textsuperscript{1,*}
\\
\bigskip
\textbf{1} Department of diagnostic and interventional Radiology, Technical University of Munich, Munich, Germany
\\
\textbf{2} Institute for Artificial Intelligence in Medicine and Healthcare, Technical University of Munich, Munich, Germany
\\
\textbf{3} OpenMined
\\
\textbf{4} Department of Computing, Imperial College London, London, United Kingdom
\\
\bigskip
\Yinyang These authors contributed equally to this work.
\\
* Corresponding author e-mail: rbraren@tum.de

\end{flushleft}
\section*{Abstract}
\noindent 
The success of deep learning in recent years has arguably been driven by the availability of large datasets for training powerful predictive algorithms. In medical applications, the sensitive nature of the data limits the collection and exchange of large-scale datasets. Privacy-preserving and collaborative learning systems can enable the successful application  of machine learning in medicine. However, collaborative protocols such as federated learning require the frequent transfer of parameter updates over a network. To enable the deployment of such protocols to a wide range of systems with varying computational performance, efficient deep learning architectures for resource-constrained environments are required.
\par 
Here we present \textit{MoNet}, a small, highly optimized neural-network-based segmentation algorithm leveraging efficient multi-scale image features. \textit{MoNet} is a shallow, \textit{U-Net}-like architecture based on repeated, dilated convolutions with decreasing dilation rates. We apply and test our architecture on the challenging clinical task of  pancreatic segmentation in computed tomography images. We assess our model's segmentation performance and demonstrate that it provides superior out-of-sample generalization performance, outperforming larger architectures, while utilizing significantly fewer parameters. We furthermore confirm the suitability of our architecture for federated learning applications by demonstrating a substantial reduction in serialized model storage requirement as a surrogate for network data transfer. Finally, we evaluate \textit{MoNet}'s inference latency on the central processing unit (CPU) to determine its utility in environments without access to graphics processing units.\par
Our implementation is publicly available as free and open-source software.

\newpage

\section*{Introduction}
Access to large collections of data remains one of the key challenges in successfully applying machine learning to many problems in medicine. Common machine learning datasets, such as ImageNet\cite{imagenet} with $>1$ million images, are much larger than their counterparts used in medical studies. Even large recent studies\cite{mckinney2020international, bora2020predicting} use datasets significantly smaller than ImageNet and orders of magnitude smaller than the datasets used to train state-of-the-art language models\cite{devlin2018bert}.
Furthermore, current medical studies often source data from only few institutions, thus preventing the training of representative and unbiased models, suitable for application in a broad variety of patient collectives\cite{obermeyer2019dissecting}. Algorithms trained on single-institutional data have recently been shown to cause generalization challenges to out-of-sample data\cite{zhang2020generalizing}. One of the main hindrances to large-scale, multi-institutional medical data collection, which could address this challenge, is the strict regulation of patient data, preventing its exchange and mandating the development of decentralized learning systems\cite{winter2019ContextIntegrity}.\par
Federated machine learning\cite{federated2015} allows for collaborative training of algorithms on data from different hospitals (\textit{data silos}) or edge devices (such as wearable health sensors or mobile phones) without the need for central aggregation of said data. In federated learning, a model is trained in a distributed fashion. Individual models are trained locally on data which never leaves a participating site (\textit{node}), and only parameter updates are sent via the network to be aggregated by the coordinating node (\textit{hub-and-spoke topology}).  Federated learning enhanced by privacy-preserving techniques\cite{kaissis2020secure} such as differential privacy\cite{dwork2014algorithmic} holds the promise of secure, large-scale machine learning on confidential, medical data.\par
The utilization of federated learning techniques on the largest possible number of institutions and patients from a diverse geographic, demographic and socio-economic background will require the development of systems suitable for execution on a broad range of hardware including mobile devices and systems without graphics processing units, which may be too expensive for deployment e.g. in the developing world. A further key component of this democratization is the improvement of system efficiency, as federated learning requires the frequent transfer of parameter updates over a network. Previous work\cite{li2020federatedChallenges} has mainly focused on improving communication efficiency by compression of parameter updates or sophisticated update aggregation methods\cite{konevcny2016federated, bonawitz2019FedLearningAtScale}. However, the targeted design of small and efficient neural network architectures, targeting both above mentioned points has so far remained under-explored. To the contrary, deep learning has focused on expanding model size with current models possessing upwards of one billion parameters \cite{brown2020languagegpt3}.
\par
Here, we introduce \textit{MoNet}, a very small, shallow, \textit{U-Net}-derived semantic segmentation architecture based on efficient multi-scale feature extraction using repeated decreasingly dilated convolution (RDDC) layers with two global down-sampling operations and a total of 403,556 parameters. We showcase our architecture's performance on the challenging task of pancreatic segmentation and demonstrate substantial efficiency gains and segmentation performance competitive with much larger models.

%%%%%%%%%%%%%%%%%%%%%%%%%%%%%%%%%%%%%%%%%%%%%%%%%%%%%%%%%%%%%%%%%%%%%%%%%%%%%%%%
\section*{Methods}
\subsection*{Training, validation and independent testing datasets}
All neural network architectures presented in this work were trained on the pancreas dataset from the Medical Segmentation Decathlon (MSD) \cite{simpson2019large}. A random, consistent 70\%/30\% training-validation split was employed. Hence, 196 abdominal CT scans of the portal-venous contrast agent enhancement phase were used for training and 85 scans for validation. For processing, images were bilinearly down-sampled to $256 \times 256$, and the pancreas and tumor labels were merged yielding a binary segmentation task. To assess out-of-sample generalization performance, independent validation of the architectures was performed on an unseen, clinical PDAC dataset consisting of 85 abdominal CT scans in the portal-venous phase collected at our institution. All clinical data were collected according to Good Clinical Practice and in consent with the Declaration of Helsinki. The use of imaging data was approved by the institutional ethics committee (Ethikkommission der Fakultät für Medizin der Technischen Universität München, protocol number 180/17S, May 9th 2017) and the requirement for informed written consent was waived. The pancreas including the tumor was manually segmented by a third-year radiology resident, then checked and corrected as necessary by a sub-specialized abdominal radiologist. An exemplary ground truth label mask superimposed on a CT slice from the training set is shown in Figure 1.

    \begin{figure}[H]
        \centering
        \includegraphics[scale=0.1]{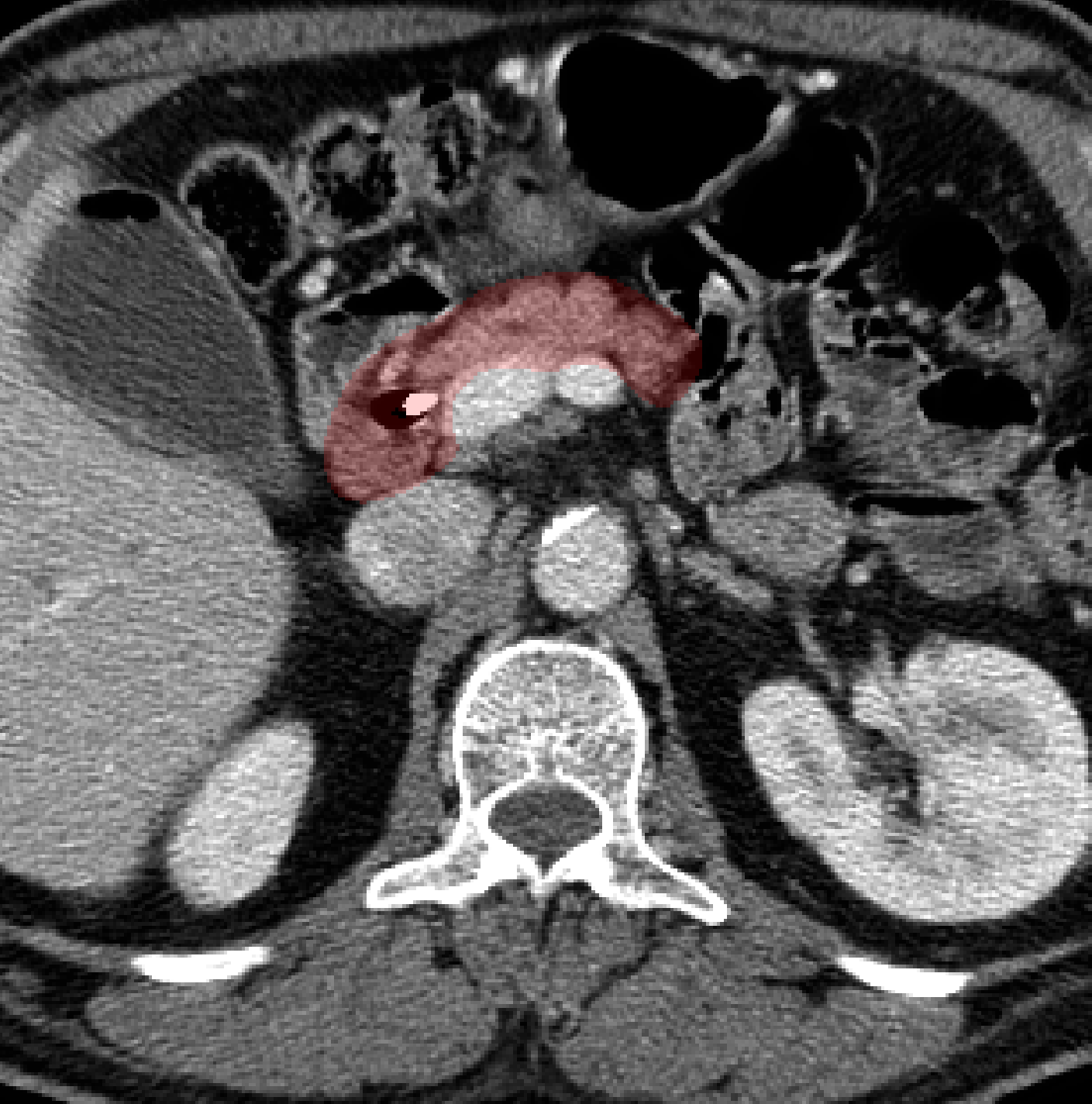}
        \caption{Axial slice of a ground truth pancreas segmentation in an abdominal CT scan (MSD), cropped to show detail of surrounding tissues}
        \label{fig:pancreas}
    \end{figure}

\subsection*{Network architecture}
The architecture of \textit{MoNet} is depicted in Fig. 2. In brief, 2-dimensional input tensors of shape $B\times256\times256\times1$, with $B$ denoting the batch size, are progressively down-sampled across the encoder branch of the network using  convolutions with a stride length of $2$, resulting in an $X\times Y$ resolution of $64\times64$ in the \textit{bottleneck} segment of the network. The resulting feature maps are then progressively up-sampled by transposed convolution (\textit{deconvolution}) in the decoder branch resulting in output masks of identical dimensions as the input. Each (de-)convolution block consists of a 3x3 convolutional layer followed by batch normalization and an \textit{exponential linear unit} (ELU) activation. At every stage in the \textit{U-Net}-like architecture, the convolution blocks are followed by a \textit{repeated decreasingly dilated convolution} (RDDC) block (Fig. 3), consisting of four successive convolutional blocks as described above, but employing dilated convolutions \cite{holschneider1990real} with a decreasing dilation rate (4, 3, 2, 1, respectively). This feature extraction strategy has been shown to perform well for small objects\cite{hamaguchi2018effectiveDilatedConvs}. Each convolutional block within a \textit{RDDC} block is followed by a spatial dropout layer\cite{tompson2015efficient}. Finally, residual-type longitudinal (short) connections are employed within each RDDC block and transverse (long) skip connections are employed between the encoder and the decoder branch to assist signal and gradient flow as originally described in \cite{ronneberger2015u, resnetHe2016}.

    \begin{figure*}[h]
    \centering
    \includegraphics[width=1.\textwidth]{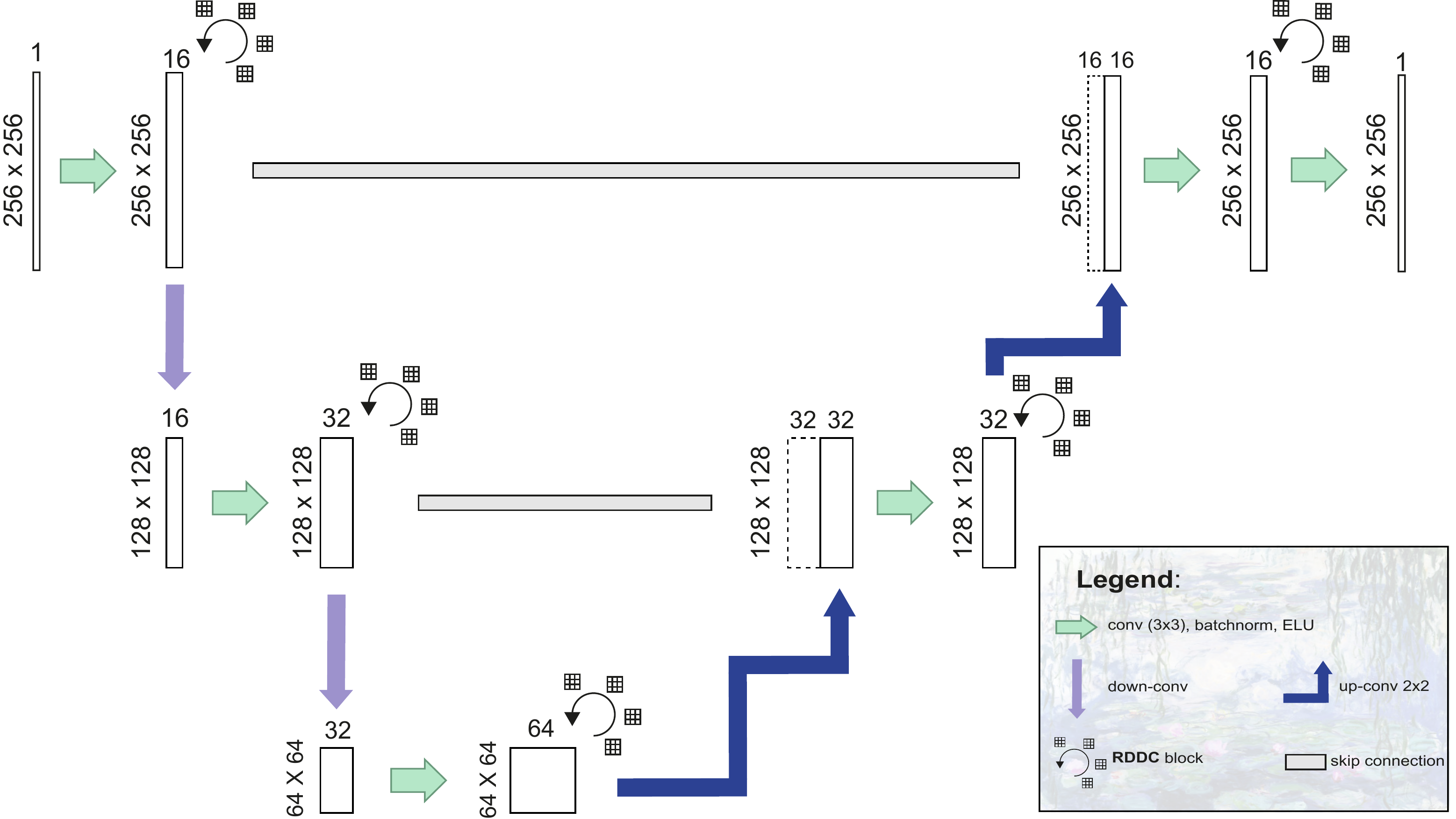}
    \caption{Schematic representation of the \textit{MoNet} architecture.}
    \label{fig:monet_architecture}
    \end{figure*}

    \begin{figure}[h]
        \centering
        \includegraphics[width=1.\textwidth]{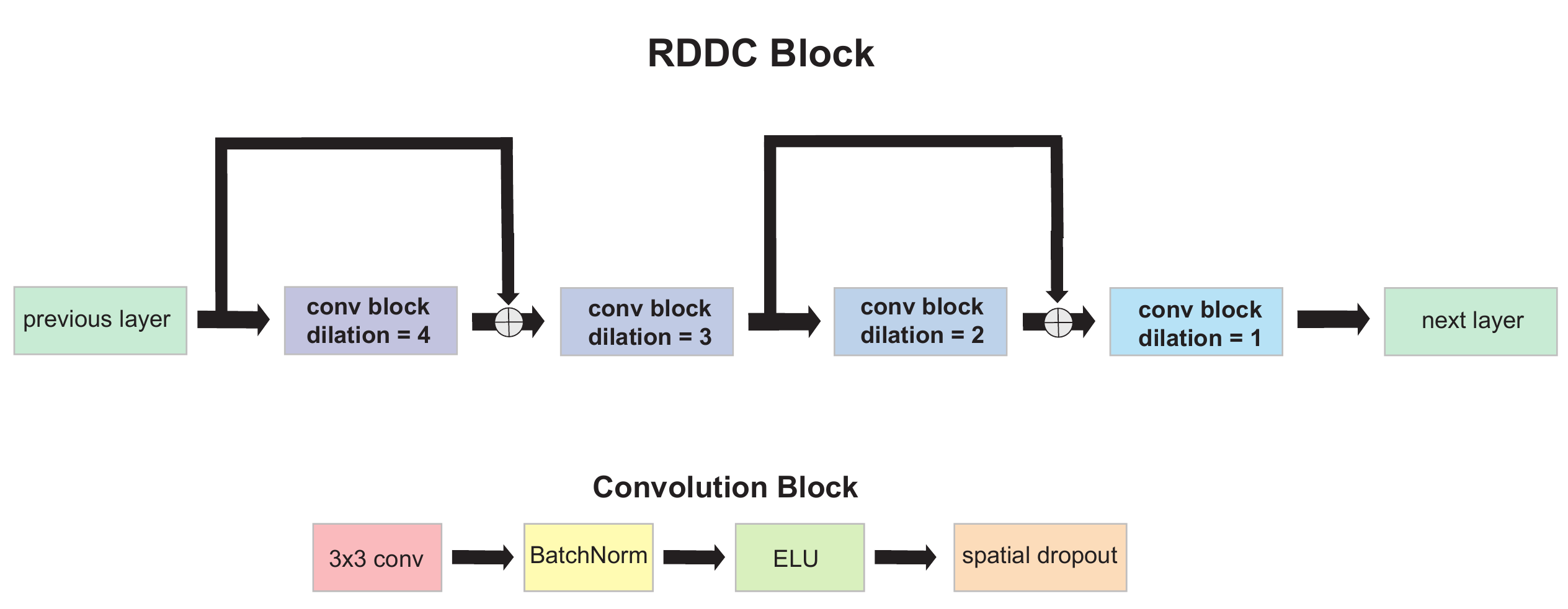}
        \caption{Schematic representation diagram of a RDDC block (top) and the constituent convolutional blocks (bottom).}
        \label{fig:rddc_block}
    \end{figure}

\subsection*{Model training}
All architectures were trained to convergence using the \textit{Nesterov-Adam} optimizer\cite{NAdamDozat2016} with an initial learning rate of $5\times10^{-4}$ and learning rate decay by a factor 10 upon validation loss stagnation for $\geq$ 2 epochs. Weights were initialized using uniform He-initialization\cite{he2016deep} and the  Dice loss\cite{milletari2016v} was used to train all networks. Data augmentation was used in the form of random rotations up to 10 degrees, random zoom ($\pm 0.25$) and random pixel shifts of a maximum magnitude of 0.2 of the image height/width. All architectures were trained to segment the entire pancreas including the tumor. This approach is owing to the fact that the exact delineation of the tumor border is often times infeasible and supported by literature findings noting the importance of the peritumoral tissue in PDAC \cite{Bauer2017, Fukushima2005, Infante2007} and in other tumor entities \cite{Sun2020}.

\subsection*{Performance Assessment}
We compared \textit{MoNet}'s out-of-sample generalization performance to the following three \textit{U-Net} baselines: 
\begin{itemize}
  \item original \textit{U-Net}\cite{ronneberger2015u}, 64 base filters (\textit{U-Net-64})
  \item original \textit{U-net}\cite{ronneberger2015u}, 16 base filters (\textit{U-net-16})
  \item Attention-gated \textit{U-Net}\cite{oktay2018attention}, 2D, 64 base filters (\textit{Attention U-Net})
\end{itemize}

\noindent For all run-time comparisons, repeated testing was performed under identical circumstances (no concurrent network traffic, all non-essential operating system processes suspended, identical CPU power settings). %Mean inference times and Dice scores were compared using the Student's t-test with multiple testing correction.

\section*{Results}

\subsection*{Segmentation performance comparison}
\textit{MoNet} performed on par with other \textit{U-Net} variants on the validation dataset while outperforming the other \textit{U-Net} variants on the independent validation dataset. Results are summarized in Table 1.

    \begin{table}[h]
            \centering
    				\resizebox{0.9\textwidth}{!}{\begin{tabular}{c c c }
    					\hline
    						 \textbf{Architecture} & \textbf{Mean$\pm$STD Dice, MSD} & \textbf{Mean$\pm$STD Dice, IVD}\\
    					\hline
    						\textit{U-Net-64} & $0.70 \pm 0.15$ & $0.50 \pm 0.2$ \\
    						\textit{U-Net-16} & $0.67 \pm 0.14$ & $0.59 \pm 0.2$ \\
    						\textit{Attention U-Net} & $0.66 \pm  0.15$ & $0.37 \pm 0.6$\\
    						\textit{MoNet (ours)} & \boldmath{$0.74\pm0.11$}  & \boldmath{$0.70\pm0.1$}\\
    					\hline
    				\end{tabular}}
    				\caption{Comparison of \textit{MoNet} with other \textit{U-Net} variants tested on the MSD validation set and the independent validation set (IVD) (both N=85 scans).}
    			\label{results_table}
    \end{table}

\subsection*{CPU inference-time comparison}
A comparison of the time required for performing inference with 150 $256\times 256$ images on CPU (2.4GHz 8-Core Intel Core i9) was performed with identical batch size and otherwise consistent environment for \textit{U-Net}, \textit{Attention U-Net} (2D) and \textit{MoNet}. \textit{MoNet} significantly outperformed both \textit{U-Net} and \textit{Attention U-Net} with regards to inference time. Results are shown in Table 2.

    \begin{table}[H]
        \centering
            \resizebox{0.6\textwidth}{!}{\begin{tabular}{c c c}
            \hline
             \textbf{Architecture} & \textbf{Mean$\pm$STD inference times} \\
             \hline
              \textit{U-Net-64} & $45.34 \pm 1.77$\\
              \textit{U-Net-16} & \boldmath{$7.03  \pm 0.21$}\\
              \textit{Attention U-Net} & $53.30 \pm 0.53$\\
             \textit{MoNet (ours)} & $14.88\pm0.32$  \\
             \hline
        \end{tabular}}
        \caption{CPU inference time (sec) for a CT scan of 150 slices at $256\times256$ resolution, results averaged over 5 runs under identical setup.}
        \label{tab:inference_comparision}
    \end{table}

\subsection*{Serialized model size as an indicator for network traffic in federated learning}
We performed a comparison of the size taken up by the weights of \textit{MoNet} and the other \textit{U-Net} like architectures. Federated learning requires the frequent transfer of parameter updates over a network, hence the serialized model size of a given architecture can serve as an estimate of the amount of network traffic generated when deployed in a federated learning application. \textit{MoNet} with its small number of parameters is significantly smaller in size than \textit{U-Net-16} and an order of magnitude smaller than \textit{U-Net-64} and \textit{Attention U-Net}. Results are shown in Table 3.

    \begin{table}[ht]
    				\center
    				\resizebox{0.7\textwidth}{!}{\begin{tabular}{c c c}
    					\hline
    						 \textbf{Architecture} & \textbf{Parameter count} & \textbf{Size in memory} \\
    					\hline
    						\textit{U-Net-64}& $31,054,145$ & 118.6 MB\\
    						\textit{U-Net-16}& $1,946,705$ & 7.6 MB\\
    						\textit{Attention U-Net}& $31,753,349$ & 121.3 MB\\
    						\textit{MoNet (ours)}	& $403,556$ & \textbf{1.8 MB}\\
    					\hline
    				\end{tabular}}
    				\caption{Comparison of storage space occupied by \textit{MoNet} and other \textit{U-Net} variants.}
    			\label{size:comparison}
    \end{table}

\newpage

%%%%%%%%%%%%%%%%%%%%%%%%%%%%%%%%%%%%%%%%%%%%%%%%%%%%%%%%%%%%%%%%%%%%%%%%%%%%%%%%
\section*{Discussion}
We here present an efficient, high-performance \textit{U-Net}-like segmentation algorithm and show a substantial reduction in parameter count and generated network traffic in a federated learning application (indicated by serialized model size). Compared to \textit{U-Net-64} and \textit{Attention U-Net}, our method achieves a substantial inference latency reduction on CPU hardware while exceeding the segmentation performance of all evaluated algorithms. We thus believe our architecture to be a promising candidate for utilization in collaborative medical imaging workflows.  \par
We chose the task of pancreatic segmentation due to the poor prognosis and increasing incidence of PDAC \cite{ACS2020, collisson2019molecular}, which mandate the development of enhanced diagnosis and treatment strategies. Our recent findings suggest that quantitative image analysis can identify molecular subtypes related to different response to chemotherapeutic drugs \cite{Kaissis2020a} or predict patient survival \cite{Kaissis2020}. Automated region-of-interest definition increases the reliability and validity of such findings, and offers substantial time savings compared to manual expert-based segmentation. However, the success of automated pancreatic segmentation algorithms is constrained by the organ's poor differentiability from adjacent structures of similar attenuation, variability in position and fat content and alterations due to pathology such as tumor or inflammation.
Existent work in deep learning-assisted semantic segmentation of medical images and the pancreas in particular has focused on expanding previously available architectures such as the \textit{U-Net}\cite{ronneberger2015u} into the three-dimensional context \cite{milletari2016v} or on improving segmentation results by incorporating attention mechanisms into the architecture\cite{oktay2018attention}. Other approaches have used complex ensembles of 2D and 3D models to extract the maximum amount of information in the CT images \cite{isensee2018nnu}.
All these modifications however result in a further increase in the (already substantial) computational requirements of these architectures, rendering such \textit{U-Net} derivatives impractical for the utilization in the above-mentioned decentralized learning applications.\par
Our method enables competitive segmentation performance with the state-of-the-art while offering substantial efficiency gains through the utilization of higher resolution feature maps in the \textit{bottleneck} section of the network. Recent work on semantic segmentation provides evidence in favor of architectures performing image feature extraction at multiple scales by utilizing dilated convolutions instead of relying merely on the scale-decreasing backbones employed in traditional fully convolutional architectures \cite{spineNet2020, chen2017DeepLabV3,hamaguchi2018effectiveDilatedConvs, yu2015dilatedconvsMultiScale}. Our work corroborates this notion, since multi-scale feature extraction combined with larger receptive fields at the same hierarchical level seem to capture both more robust and higher quality features compared to the fixed kernel size design encountered in \textit{U-Net}-like architectures. Moreover, architectures with several down-sampling operations and/or many filters such as the \textit{U-Net} (with 4 down-sampling stages) cannot leverage the large number of parameters sufficiently well to warrant their utilization at least in medical imaging tasks, typically characterized by small segmentation targets (such as the pancreas or small tumors).\par
Our results indicate that \textit{MoNet} extracts more robust features that generalize better to out-of-sample data than the compared methods, as shown by \textit{MoNet}'s performance on the independent validation set. The poor performance of the 64 filter \textit{U-Net} and \textit{Attention U-Net} in the out-of-sample generalization challenge could potentially be caused by the overparameterization of these architectures, making them prone to overfitting the data-generating distribution of the training data, while the two smaller models(\textit{U-Net-16} and \textit{MoNet}) tested seemed to generalize better to the out-of-sample data, supporting this hypothesis.\par
Our work is not without limitations. The generalizability of our findings should be confirmed using larger, multi-institutional training and validation sets. Furthermore, we only compared our algorithm against models based on the use of a single 2D-\textit{U-Net}-style network. Algorithms such as \textit{nnU-Net} \cite{isensee2018nnu} based on \textit{U-Net} ensembles offer superior performance, however at the expense of extremely high computational and post-processing requirements and thus much slower inference times (especially on CPU). Furthermore implementing and establishing a real-world federated learning application for pancreatic segmentation was out of scope for this study and will be addressed in future work.

\section*{Conclusion}
In conclusion, we propose an optimized semantic segmentation algorithm with small size and low inference latency, particularly suited for decentralized applications such as federated learning.
Our work can benefit both, radiological research and clinical translation of artificial intelligence workflows in medical imaging by providing consistent, high-quality segmentation for machine learning tasks.

\section*{Source code and data availability}
Source code for \textit{MoNet} based on TensorFlow is available at \url{https://github.com/TUM-AIMED/MoNet}. The training datasets are available from \url{http://medicaldecathlon.com/}. The independent test set data contains confidential patient information and cannot be shared publicly.

\section*{Acknowledgments}
Authors wish to thank Alexander Ziller and Nicolas Remerscheid for their scientific input, Novi Quadrianto for his support in supervising the bachelor thesis that this work stems from and Karl Schulze for the excellent graphic design of the figures.

% Use the PLoS provided BiBTeX style
\bibliographystyle{plos2015}
\bibliography{bibliography.bib}

\end{document}